\documentclass[12pt]{iopart}
\newcommand{\dSum}{\displaystyle\sum}
\newcommand{\dProd}{\displaystyle\prod}
\usepackage{graphicx}
\usepackage{subfigure}
\begin{document}

\title[Bulk properties from small simulations]{Extracting bulk properties of self-assembling systems from small simulations}

\author{Thomas E. Ouldridge$^1$, Ard A. Louis$^1$ and Jonathan P. K. Doye$^2$}

\address{$^1$ Rudolf Peierls Centre for Theoretical Physics,
 University of Oxford, 1 Keble Road, Oxford, OX1 3NP, United Kingdom}
\address{$^2$ Physical and Theoretical Chemistry Laboratory,
  Department of Chemistry, University of Oxford, South Parks Road,
  Oxford, OX1 3QZ, United Kingdom}
\begin{abstract}
For systems that self assemble into finite-sized objects, 
it is sometimes convenient to compute the thermodynamics 
for a small system where a single assembly can form. 
However, we show that in the canonical 
ensemble the use of small systems can lead to 
significant finite-size effects due to the suppression of concentration 
fluctuations. We introduce
methods to estimate the bulk-yields from simulations of small systems and
to follow the convergence of yields with system size, under the assumptions
that the various species behave ideally. We also propose an extension to the
umbrella sampling technique that allows the formation of multiple finite-sized objects.

\end{abstract}

\section{Introduction}

Self-assembly of monomer units into structures with a characteristic finite size is one of the central themes of soft matter physics, examples including the formation of DNA duplexes, protein complexes, virus capsids and spherical micelles.
Of course, an important aspect of such systems is their thermodynamics, 
particularly in how it affects the assembly dynamics. However, how to best
obtain the equilibrium thermodynamics from simulations of assembling systems
is an important open question and one that has become more pressing, 
as increased computer power and the availability of coarse-grained models has 
made such simulations more feasible.

If self-assembly occurs relatively readily, the best approach is probably 
to directly simulate a sufficiently large system where a number of 
the assembled structures can simultaneously form. After equilibration has taken place, statistics can then be extracted for the frequency with which clusters of
various sizes are observed. For example, such an approach has been used 
to characterize the thermodynamics of micellization \cite{Floriano99,Panagiotopoulos02}.

However, it is not uncommon that such direct approaches are problematic, 
because the presence of significant free energy barriers to assembly 
makes equilibrium hard to achieve. 
These difficult cases typically involve either the association of a relatively 
large number of simple monomers or a small number of complicated monomers.
Examples of the former include the assembly of coarse-grained models of proteins
and patchy colloids into virus capsid-like objects 
\cite{Hagan2006,Rapaport2008,Wilber2007,Nguyen2007,Wilber09,Wilber09b}, while 
examples of the latter include the formation of DNA 
duplexes \cite{Sambriski2008,Sambriski2009,Ouldridge2009}, 
or the assembly of larger DNA nanostructures \cite{Ouldridge2009}.

In these cases, an obvious way to facilitate assembly would be to use rare-event
techniques, such as umbrella sampling \cite{Torrie1977,Frenkel2001}, 
and which usually require an order
parameter that can characterize the transition to the assembled state. 
Usually it is relatively straightforward
to conceive of order parameters that can be used to describe the formation 
of a single target structure, e.g.\ the number of correct base pairs for 
assembly of a DNA duplex \cite{Sambriski2008,Sambriski2009,Ouldridge2009},
or the number of particles in the largest cluster for the assembly 
of capsid-like objects \cite{Wilber2007,Nguyen2007,Wilber09,Wilber09b} 
or micelles \cite{Pool2005}. 
However, it is less clear how to develop an order parameter 
that can be used to drive the formation of multiple target structures, 
and we are not aware of any such studies in the literature that achieve this.

Therefore, an appealing approach to obtain the equilibrium thermodynamics
would be to simulate the formation of a single assembled structure, using 
say umbrella sampling, and if the self-assembly is monodisperse, 
i.e.\ the assembled structure has a specific size, to perform the simulation 
in the canonical ensemble with exactly the right number of particles to form one
complete structure. But what would be the errors with such an approach? 
Firstly, interactions between the assembled structures are neglected. 
However, this is often likely to be a relatively good approximation, because 
the interactions between assembled structures are likely to be mainly associated
with excluded volume---any attractions are likely to be weak compared to the
forces associated with the assembly itself--and assembly often occurs at 
relatively low concentrations. 

The second potential source of error is finite-size effects, and these
are the focus of the present paper. In particular, we will show that these
finite-size errors in canonical ensemble simulations can be 
significant\footnote{We note there will also be finite-size errors associated 
with grand canonical simulations if the system is {\it restricted} to form a single assembled structure.},
but also how they can be corrected under the assumption that species behave 
ideally\footnote{We note that in Ref.\ \cite{Wilber2007} we did not apply these 
corrections when comparing the thermodynamic yield from single target 
simulations to the bulk dynamical yield. Nor is there any mention of corrections
being applied to the DNA duplex results in Refs.\ \cite{Sambriski2008,Sambriski2009}}, which as we mentioned above is often
a good approximation. We will also examine how the assembly yields converge 
towards the bulk values as the system size is increased, as well as 
suggesting an extension to the umbrella sampling scheme which facilitates the study of multiple targets of self-assembly.

We should note that specific methods have been developed to calculate the equilibrium thermodynamics of heterodimer formation in the protein-ligand binding 
literature \cite{Gilson2007}. 
These aim to estimate the partition function of bound and separated molecules directly, and typical methods include calculating a potential of mean force for a certain pathway between bound and unbound structures, incorporating the effects of overall translational degrees of freedom separately. 
These techniques, which are optimized for problems of great computational difficulty, lack the flexibility and simplicity of the approach analysed here, in which nothing need be approximated or assumed about the nature of bonding, and no pathway need be imposed. Furthermore, they do not generalize well to multicomponent assembly.

\section{Dimer formation in the canonical ensemble}
\subsection{Heterodimer formation}
\label{heterodimer thermo}
\begin{figure}
\begin{center}
\includegraphics[width=7.2cm]{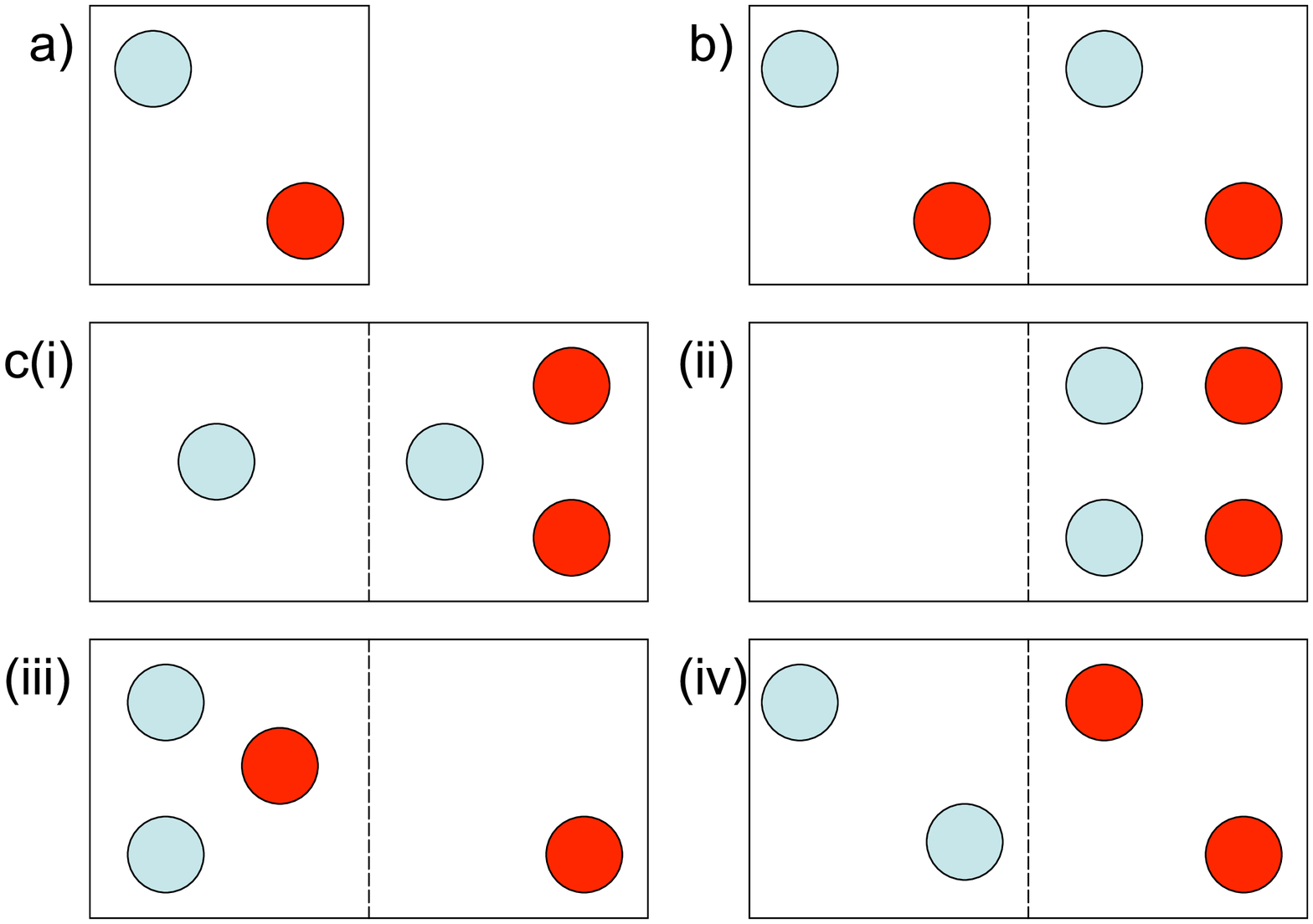}
\end{center}
\caption{\footnotesize{a) Shows two different particles in a box of size $v$ as discussed. We now imagine doubling the volume to $2v$ whilst doubling the number of particles, therefore maintaining the average concentration, as shown in b). For the purposes of analysis, we can split our new volume in half with a dashed line as indicated. All macrostates with one of each of A and B on either side of the dashed line, such as b), will provide the same statistics as the original system. Macrostates such as those shown in c), however, will have different statistics: for instance, c(iv) necessarily has a binding fraction of zero.}}
\label{fluctuations}
\end{figure}

We illustrate the physical cause of finite size statistical corrections by considering heterodimer formation---such a system may correspond to protein binding or DNA hybridization \cite{Sambriski2008, Sambriski2009}. We consider a simulation in a periodic cell of volume $v$, containing one monomer of type `A' and one of type `B'. Assuming we have a criterion for defining a subset of states as `bound', our simulation will estimate the relative probability with which bound (AB) and unbound (A,B) states are observed in such a system:
\begin{equation}
\Phi = \frac{{\rm probability \rm (AB)}}{{\rm probability (A,B)}},
\label{relative partition function}
\end{equation} 
with the fraction of bound pairs given by:
 \begin{equation}
 f_1 = \Phi/(1+\Phi).
 \label{fraction bound 1}
 \end{equation}
Naively, we might hope that $f_1 = f_\infty$, the bulk equilibrium bonding fraction at the same temperature and concentration. Unfortunately, this is not the case, as although we match the average concentration of a bulk system we do not match concentration fluctuations (as shown in figure \ref{fluctuations}) in a small volume of that system.

We can apply corrections using simple thermodynamic arguments, if we assume that interactions between all particles which are not in a dimer state are negligible. We consider a system of volume $Dv$, with $D$ an integer, with the same average concentration as the system with $D=1$. We define:
\begin{itemize}
	\item $Z_{\rm AB}$ and $Z_{\rm A,B}$ as the partial partition functions of the $D=1$ system when confined to the relevant subset of states. For future convenience, we define these quantities using {\em distinguishable statistics}, although it does not matter at this stage. We note that $\Phi = Z_{\rm AB}/ Z_{\rm A,B}$.
	\item $N$ as the total number of particles of type A or B (here $N=D$).
	\item $N_i$ as the number of molecules of species $i$ (in this case $i$ is A, B or AB).
	\item $q_i$ as the single particle partition function for species $i$, in the volume $Dv$, with the internal degrees of freedom treated using {\em indistinguishable statistics}.
	\item $\mu_i$ as the chemical potential of species $i$.
\end{itemize}
The $\mu_i$ are given by a standard result of statistical mechanics:
\begin{equation}
\mu_{\rm i} = -{\rm k_B}T \frac{\partial}{\partial N_{\rm i}} \ln \left ( \frac{q_{\rm i}^{N_{\rm i}}}{N_{\rm i}!} \right ) \approx  -{\rm k_B}T \ln \left ( \frac{q_{\rm i}}{N_{\rm i}} \right ),
\label{mu_i}
\end{equation}
where the approximation becomes an equality in the thermodynamic limit. In this limit, we can use the standard equilibrium result $\sum_i \nu_i \mu_i = 0$, where $\nu_i$ are the stoichiometric coefficients of the species in the reaction, finding:
\begin{equation}
\frac{N_{\rm AB}}{N_{\rm A}N_{\rm B}} = \frac{q_{\rm AB}} {q_{\rm A} q_{\rm B}}.
\label{thermodynamics prediction}
\end{equation}
As each $q_{\rm i}$ scales with the volume of the system, we have:
\begin{equation}
q_{\rm AB} = D Z_{\rm AB} ,
\label{association of q with Phi 1}
\end{equation}
\begin{equation}
q_{\rm A} q_{\rm B} = D^2 Z_{\rm A,B},
\label{association of q with Phi 2}
\end{equation}
which gives (using $D=N$):
\begin{equation}
\frac{[{\rm AB}]}{{[\rm A}][{\rm B}]} = \frac{v Z_{\rm AB}} {Z_{\rm A, B}} = v \Phi = K^{\rm eq}_{\rm A,B}.
\label{conc eqn}
\end{equation}
We note that the quantity $Z_{\rm AB} / Z_{\rm A, B}$ is that which is generally directly estimated in protein/ligand binding studies \cite{Gilson2007}; this is then multiplied by a reference concentration to give the equilibrium constant.

\begin{figure}
\begin{center}
\includegraphics[width=5.2cm,angle=-90]{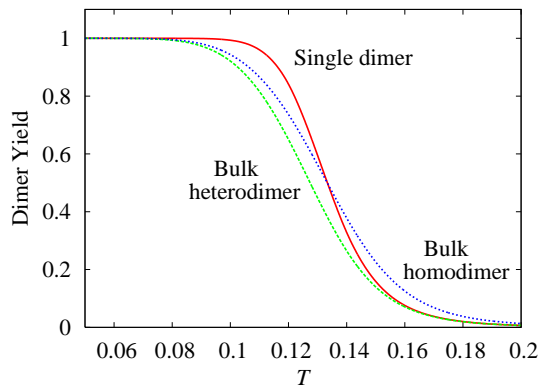}
\end{center}
\caption{\footnotesize Dimer yield for a system described by a two-state model $Z_2 / Z_{1,1} = \exp(-\Delta E / T + \Delta S)$, with $\Delta E = 2$ and $\Delta S = 15$ in reduced units, with the values chosen for illustrative convenience. Plotted are the yield for a two-particle system and the bulk values at the same average concentration for homodimers and heterodimers.}
\label{transition}
\end{figure}

Substituting (\ref{relative partition function}), (\ref{association of q with Phi 1}) and  (\ref{association of q with Phi 2}) into (\ref{thermodynamics prediction}) yields:
\begin{equation}
\frac{f_\infty}{(1-f_\infty)^2} = \Phi
\label{Phi equation}
\end{equation} 
\begin{equation}
=> f_\infty = \left(1+\frac{1}{2\Phi}\right) - \sqrt{\left(1+\frac{1}{2\Phi}\right)^2-1}.
\label{f infty} 
\end{equation}
In this case, $f_\infty < f_1$ for all values of $\Phi$, as is illustrated for a model dimer-forming system in figure \ref{transition}. It is also noticeable that the transition is wider for the bulk system.
The physical causes of these two effects will be discussed at the end of section \ref{Homodimer formation}.

\subsection{Heterodimer convergence}
\label{heterodimer microscopic}
It is useful to consider how the bonding fraction converges to the bulk result as the system size is increased from one cluster to the thermodynamic limit. We consider a system of volume $Dv$, calculating the fraction of dimers ($f_D$) as a function of $\Phi$, again neglecting interactions except dimer formation. 

Consider the macrostate with $b$ dimers formed (out of a possible $D$). The partition functions of individual monomers and dimers scale with the size of the system ($Dv$), and the partition function of the system is the product of the individual partition functions together with combinatorial factors. Using $Z_{AB}$ and $Z_{A,B}$ as defined before, the partition function of a macrostate with b dimers (using distinguishable statistics) is given by:
\begin{equation}
Z_b(D) = \frac{(D Z_{AB})^b (D^2 Z_{A,B})^{D-b}}{b!} \left (\frac{D!}{(D-b)!} \right )^2,
\label{Zb}
\end{equation}
in which the combinatorial factor is obtained from the total number of permutations of A and B ($D!^2$) divided by the permutations which exchange monomers for monomers ($(D-b!)^2)$) or dimers for dimers ($b!$). We divide by $D!^2$ to make our statistics indistinguishable. We find $f_D$ in the usual way, using equation (\ref{relative partition function}) to simplify:

\begin{equation}
f_D = \frac{\sum_{b=1}^D b \left (\frac{\Phi}{D}\right )^b  \left (\frac{1}{(D-b)!} \right )^2 \frac{1}{b!}}{\sum_{b=0}^D D \left (\frac{\Phi}{D}\right )^b  \left (\frac{1}{(D-b)!} \right )^2 \frac{1}{b!}} = \frac{\sum_{b=0}^D b Z'_b}{\sum_{b=0}^D D Z'_b}.
\label{fN}
\end{equation}

\begin{figure}
\begin{center}
\includegraphics[width=7.2cm]{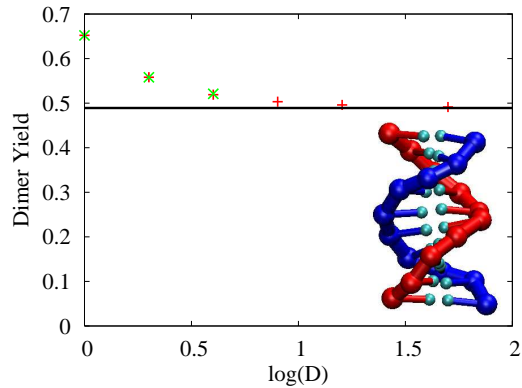}
\end{center}
\caption{\footnotesize (a) Heterodimer yield as a function of system size $D$, with average concentration fixed. 
The  `x' symbols indicate results from simulations of our DNA model
that were capable of forming $D$ DNA duplexes 5 base pairs long, 
and the `+' are the predictions of equation (\ref{fN}) 
with $\Phi$ chosen to reproduce the $D=1$ DNA result. 
The solid line indicates $f_\infty$. 
The image in the bottom right shows a duplex of 10 bases as represented by our coarse-grained DNA model. 
The simulations of the model were performed in the canonical ensemble using the Monte Carlo algorithm of Whitelam and Geissler \cite{Whitelam2007}. 
}
\label{f_D_DNA}
\end{figure}

Plotting $f_D$ against $D$ for $\Phi = 1.875$ (figure \ref{f_D_DNA}(a)) we find that the bonding fraction falls from 0.652 to a large $D$ limit of 0.489, and behaves similarly for other values of $\Phi$. 
We can formally find this limit by noting that for any value of $\Phi$, $Z'_b$ is sharply peaked about its maximum $b_{\rm mode}$ for large $D$. This allows us to make the saddle point approximation, whereby we assume that $Z'_b$ is Gaussian and therefore that $f_\infty = b_{\rm mode}/D$ by symmetry. Maximizing $\ln Z'_b$, and employing Stirling's approximation yields:
\begin{equation}
\frac{{\rm d} \ln Z'}{{\rm d}b} \approx \ln \left(\frac{\Phi}{D}\right) + 2 \ln(D - b) - \ln(b),
\label{dZ'/db}
\end{equation}
\begin{equation}
=> \frac{\Phi (D-b_{\rm mode})^2}{Db_{\rm mode}} = 1.
\label{dZ'/db=0}
\end{equation}
Using 
$b_{\rm mode}/D=f_\infty$,
we can see that equations (\ref{dZ'/db=0}) and (\ref{Phi equation}) are identical, as they should be.

The microscopic approach provides a simple mechanism for evaluating the accuracy of the correction scheme in certain cases. If it is possible to simulate the simultaneous formation of two or more targets, one can compare the change in dimer yield to the predictions of the microscopic approach, and then extend to the thermodynamic limit if the agreement is good. 
This is particularly useful if it is possible to consider an example with the relevant model where the self-assembly process is relatively simple. 
For example, we have recently developed a coarse-grained model of DNA \cite{Ouldridge_tw_2009} in which bases are represented by rigid nucleotides (inset in figure \ref{f_D_DNA}). 
All interactions between strands, such as base pairing and excluded volume, are truncated within distances much shorter than the typical separation of unbound strands, making the assumptions in deriving the corrections of the previous sections reasonable. Simulating duplex formation for short strands of 
about five bases in length is simple, and simulations forming several targets can be performed. 
The results are plotted in figure \ref{f_D_DNA}, showing perfect agreement with equation (\ref{fN}). 
Longer duplexes and complicated branched structures are much more challenging to simulate, meaning that only single target simulations are feasible. From the fact that the correction is successful for shorter duplexes, however, we can be confident that it will apply to longer strands when the concentration of DNA bases is similar.

\subsection{Homodimer formation}
\label{Homodimer formation}
It is instructive to consider the differences between homodimer and heterodimer corrections. For homodimers formed from two particles of type `A', we obtain the following expressions for the partition function of each particle species:
\begin{equation}
q_{\rm 2A} = \frac{D Z_{\rm 2A}}{2}.
\label{association of q homo}
\end{equation}
\begin{equation}
q_{\rm A}q_{\rm A} = D^2 Z_{\rm A,A},
\label{association of q homo 2}
\end{equation}
where the factor of two compensates for the overcounting of indistinguishable states in $Z_{\rm 2A}$. Proceeding as in section \ref{heterodimer thermo}, we obtain:
\begin{equation}
\frac{[2A]}{[A]^2}= \frac{vZ_{\rm 2A}}{2 Z_{\rm A,A}} = \frac{v \Phi}{2} = K_{\rm 2A}^{\rm eq}.
\label{thermodynamic result homo}
\end{equation}
The bound fraction in the thermodynamic limit follows as:
\begin{equation}
f_\infty = \left(1+\frac{1}{4\Phi}\right) - \sqrt{\left(1+\frac{1}{4\Phi}\right)^2-1}.
\label{f infty homo} 
\end{equation}
The behaviour of the correction is significantly different from that of heterodimers, as shown in figure \ref{transition}. In this case, 
 the midpoint of the transition is unchanged, but the width is noticeably larger in bulk than for the two-particle system,
i.e.\ $f_\infty >  f_1$ for $f_1 < \frac{1}{2}$, and $f_\infty <  f_1$ for $f_1 > \frac{1}{2}$.

\begin{figure}
\begin{center}
\includegraphics[width=7.2cm]{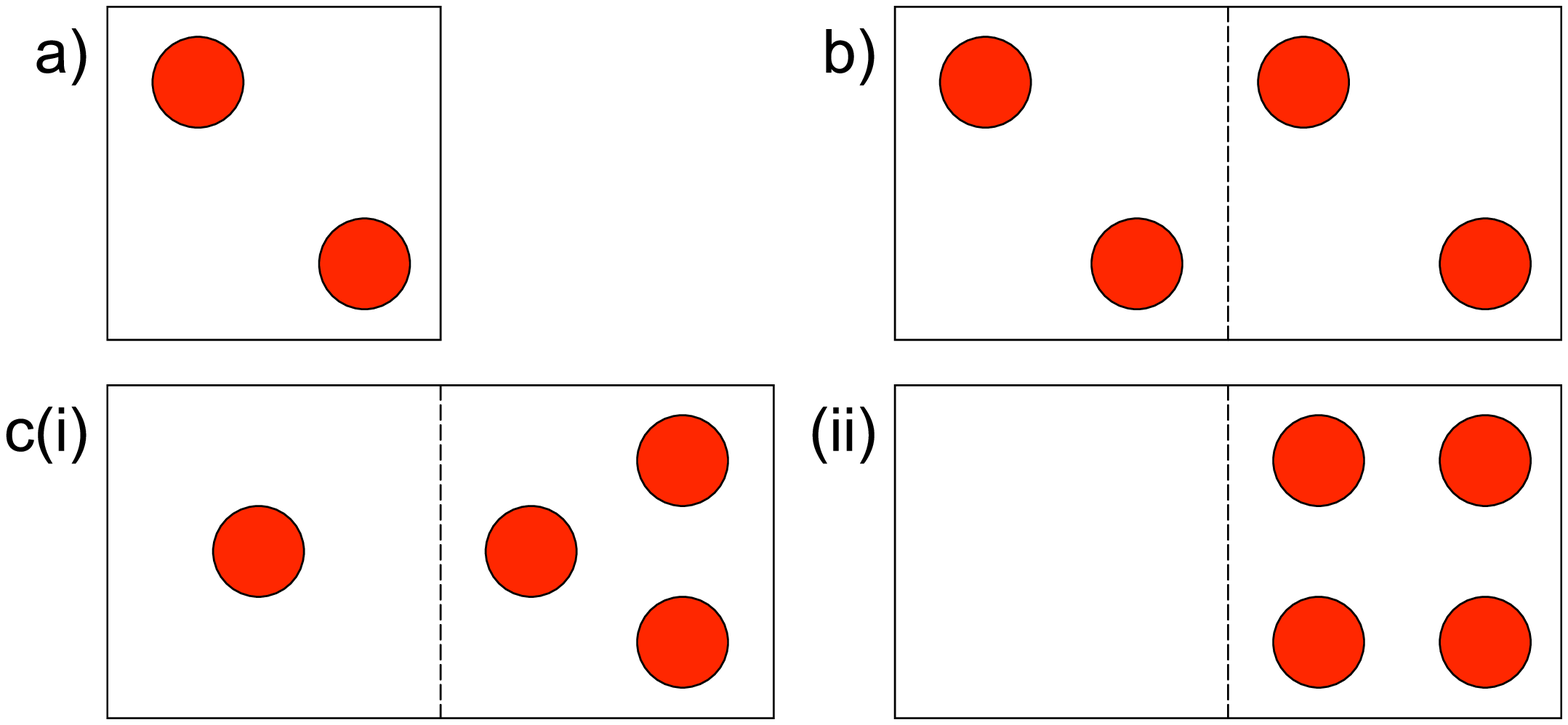}
\end{center}
\caption{\footnotesize{a) Shows two identical particles in a box of size $v$. We now imagine doubling the volume to $2v$ whilst doubling the number of particles, therefore maintaining the average concentration, as shown in b). For the purposes of analysis, we can split our new volume in half with a dashed line as indicated. All macrostates with two particles on either side of the dashed line, such as b), will provide the same statistics as the original system. Macrostates shown in c), however, will have different statistics.}}
\label{homodimer neglected states}
\end{figure}

The physical mechanism for the broadening of the transition can be understood by considering the effect of concentration fluctuations. Figure \ref{homodimer neglected states}(c) shows the states of a four-particle system which cannot be sampled in a two-particle simulation. Of these, the most probable is c(i), in which three of the particles occupy half the volume and the remainder contains only one. In this case, it is impossible to have a binding fraction of unity. A binding fraction of zero is also less likely than in the two-particle case as the three monomers occupying the `right' half of the system have a higher probability of forming one dimer than the two particles did in the original system. As a consequence, the fraction of dimers is pushed towards a half as the system grows in size, because larger concentration fluctuations are allowed which in turn favour the less probable configuration (whether dimer or monomer), leading to a broader transition in bulk.

The same argument can be applied to heterodimers, but with an important distinction. In this case, concentrations of individual species A and B can fluctuate in addition to the total concentration. 
Unlike total concentration fluctuations, fluctuations in the relative concentrations of A and B will always reduce the probability of forming dimers, 
because of configurations like that in \ref{fluctuations}c(iv) where no dimers
can be formed. 
As a consequence the heterodimer yield is lower in bulk than for a two particle system, as well as having a broader transition.

\section{Monodisperse large homoclusters}
\label{homocluster formation}
We now consider the formation of clusters of a specific size, 
a case relevant to the assembly of virus capsid-like objects \cite{Hagan2006,Rapaport2008,Wilber2007,Nguyen2007,Wilber09,Wilber09b}, and homomeric
protein complexes \cite{Villar09}.
If the formation of a single cluster is simulated in the canonical ensemble, once again, the statistics of the various cluster sizes do not directly correspond to bulk properties, but under the assumption that interactions which do not constitute bonding are negligible it is possible to extrapolate to large system sizes. For simplicity we consider clusters of identical monomers, although the analysis can be extended beyond this. Firstly, we add some definitions:
\begin{itemize}
	\item $n$ is the number of monomers needed to form the target, equal to the number of monomers simulated.
	\item $z_i$ is the partition function for species $i$, in the simulation volume $v$, with the internal degrees of freedom treated {\em distinguishably}. 
	\item $Z_{i,j,k...}$ is the partition function of a system of volume $v$ when in a state which contains one molecule of species $i$, one of species $j$ etc. This partition function is calculated using {\em distinguishable statistics}.
	\item $Z(n)$ is the total partition function of the $n$-particle system in a volume $v$, calculated using {\em distinguishable statistics}.
	\item as all monomers are identical, the `A' index will be omitted for clarity.
\end{itemize}

The thermodynamically relevant quantities are the $q_i$, because given these it is a simple task to calculate the bulk concentrations of each species using:
\begin{equation}
	\frac{N_i}{(N_1)^i} = \frac{q_i}{(q_1)^i}
	\label{mass action}
\end{equation} 
and
\begin{equation}
	\dSum_i i N_i = nD = N.
	\label{conservation}
\end{equation}

The quantities which are directly accessible from simulation are $Z_{i,j,k...}/Z$. Exactly how these can be accessed depends on how the system is sampled. A sensible choice, however, is to sample states by the largest cluster size---this neatly divides the partition function $Z$ into $n$ parts, and we label these subdivisions $\Omega_i$. We now have $n$ equations, one for each $\Omega_i$:
\begin{equation}
	\frac{\Omega_i}{Z} = \dSum_{j,k...} \frac{Z_{i,j,k...}}{Z},
	\label{n equations}
\end{equation}
where the summation over $j,k...$ is the sum over all sets of indices such that $j,k... \leq i$ and the indices sum to $n$.
We can now begin substituting for $Z_{i,j,k...}$: 
\begin{equation}
	Z_{i,j,k...} =  n! \left(\dProd_{l=i,j,k...} \frac{z_l}{l!} \right) \left(\dProd_{m}^{i} \frac{1}{C_m!} \right),
	\label{substitution}
\end{equation}
in which $C_m$ is the number of indices in the set $i,j,k...$ with the same value as m. 

We now have $n$ simultaneous equations for $z_i/Z^{i/n}$ in terms of our measured quantities $\Omega_i/Z$. In addition, these simultaneous equations have already been decoupled as each $\Omega_i/Z$ expression contains only $z_m$ with $m \leq i$, and thus finding $z_i/Z^{i/n}$ amounts to solving a polynomial of order $i$. All that remains is to find $q_i$ in terms of $z_i$. This is reasonably simple: 
\begin{equation}
	q_i = D \frac{z_i}{i!},
\end{equation}
where in dividing by $i!$ we account for the reduction in states imposed by indistinguishability. We can then obtain the right hand side of (\ref{mass action}) by:
\begin{equation}
	\frac{q_i}{(q_1)^i} = \frac{D z_i}{i! (D z_1)^i} = \frac{D z_i/Z^{i/n}}{i! (d z_1/Z^{1/n})^i},
	\label{mass action RHS}
\end{equation}
in which the right hand side is expressed in terms of the known quantities $z_i/Z^{i/n}$. We can now eliminate our arbitrary large factor $D$ by converting to concentrations (which equates to multiplying both sides by $(Dv)^{(i-1)}$), giving:
\begin{equation}
	\frac{[N_i]}{[N_1]^i} = v^{i-1}\frac{z_i/Z^{i/n}}{i! (z_1/Z^{1/n})^i}.
	\label{Mass action 2}
\end{equation}
Once again, the system of equations can be closed by conserving total monomer number:
\begin{equation}
	\dSum_i i[N_i] = n/v.
\end{equation}

\begin{figure}
\begin{center}
\includegraphics[width=5.2cm, angle=-90]{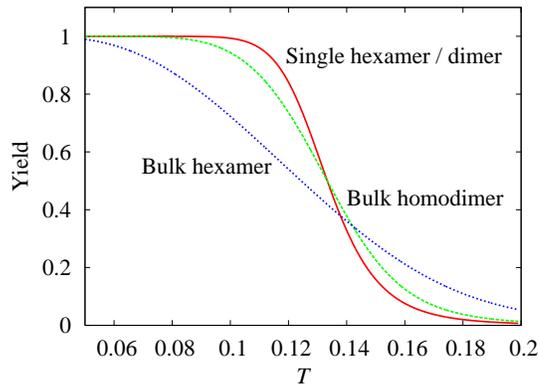}
\end{center}
\caption{\footnotesize Fractional yield of hexamers in the statistical model of equation (\ref{stat model}). Plotted are the yields for a single cluster system, for hexamers in bulk and for an equivalent extrapolation to bulk for homodimer formation.}
\label{hexamer_transition}
\end{figure}

\begin{figure}
\begin{center}
\includegraphics[width=7.2cm]{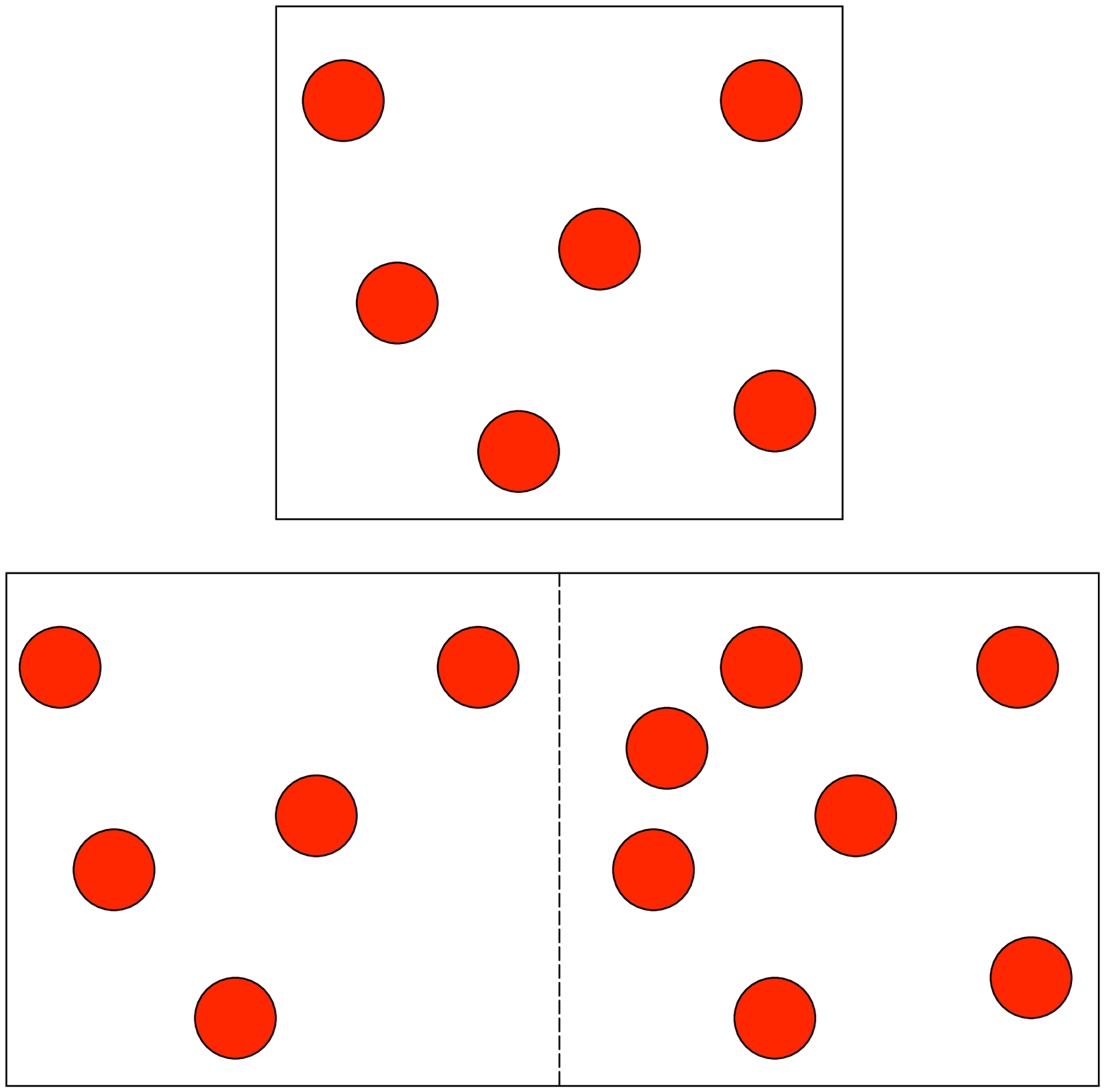}
\end{center}
\caption{\footnotesize{The top image shows six identical particles in a volume $v$. On doubling the volume, we see that only a relatively minor concentration fluctuation is required to make the formation of two hexamers impossible (and the formation of one hexamer more likely), compared to the equivalent situation for dimers. As a consequence, the broadening effect of bulk corrections increases with the size of the target structure.}}
\label{large cluster broadening}
\end{figure}

To illustrate the form of finite size corrections, we consider the artificial example of completely cooperative hexamer formation (in which we approximate clusters of intermediate size as having zero probability). The complicating effects of additional states will be discussed in section \ref{homocluster microscopic}. For comparison with section \ref{Homodimer formation}, we will assume the small system can be described by an equivalent two-state model, so that the yields of hexamers and homodimers are identical in the small simulation volume: 
\begin {equation}
Z_6 / Z_{1,1,1,1,1,1} = \exp(-\Delta E / T + \Delta S),
\label{stat model}
\end{equation}
with $\Delta E = 2$ and $\Delta S = 15$ in reduced units. The result, plotted in figure \ref{hexamer_transition}, indicates once again a much broader transition in the bulk case, this time with a slightly adjusted midpoint. 
Furthermore, this broadening is much more pronounced for hexamers than dimers. 
This trend is a general one, with larger clusters experiencing greater 
broadening due to finite-size corrections, because smaller relative 
concentration fluctuations are required to push the system towards a yield of approximately 50\% for a clustering transition involving many monomers as opposed to dimers, as illustrated in figure \ref{large cluster broadening}.

\section{Homocluster convergence}
\label{homocluster microscopic}
Many canonical simulations of self-assembly are performed using systems large enough to form several or many clusters \cite{Hagan2006,Rapaport2008,Wilber2007,Nguyen2007,Berger1998}. We apply the formalism of the previous sections to explore the convergence of cluster statistics on bulk values as system size is increased. For simplicity we restrict ourselves to a single monomer species, although the method could be extended to multiple particle types.
As in the previous section, we consider a reference system of $n$ particles in a volume $v$, where $n$ is the size of the largest cluster, and proceed using the partition functions $z_i$ defined in this volume.

\begin{figure}
\begin{center}
\includegraphics[width=7.2cm]{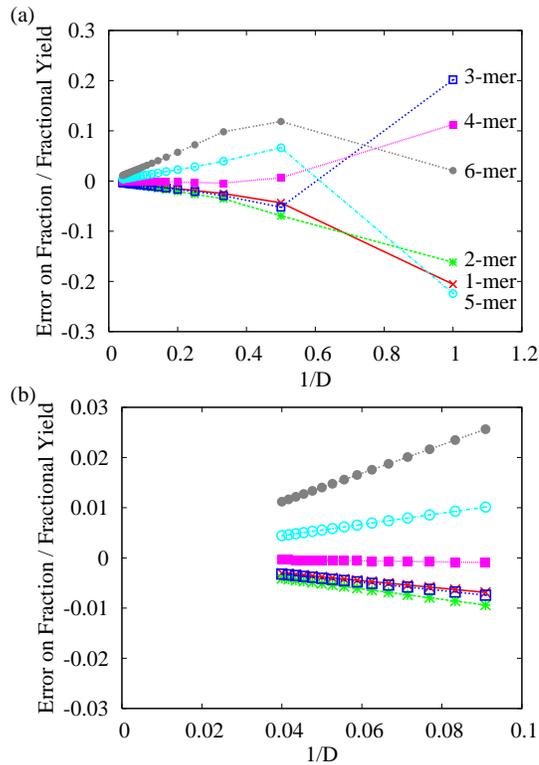}
\end{center}
\caption{\footnotesize (a) Relative fractional error on the yield of various cluster sizes as a function of $1/D$ in a system with a maximum cluster size of six. 
All clusters show convergence with $1/D$ scaling in the large $D$ regime 
(as highlighted in (b)).}
\label{1/D scaling}
\end{figure}

Let $\{\eta_i\}$ be a set of cluster sizes containing a total of $D n$ particles. The statistical weight of a state with such a set of clusters is given by:
\begin{equation}
Z_{\{\eta_i\}} = \prod_i^{n}\frac{(D z_i)^{\eta_i}}{\eta_i! (i!)^{\eta_i}},
\end{equation}
in which $\eta_i$ is the number of clusters of size $i$ in the set $\{\eta_i\}$. Defining $\psi_i(D) = z_i/(z_1^i D^{i-1})$, we obtain an expression for the fractional yield of a cluster of size $c$ in a system of size $D$:
\begin{equation}
f_c(D) =  \frac{c \sum_{\{\eta_i\}} \eta_{c} \prod_i^{n}\frac{\psi_i^{\eta_i}}{\eta_i! (i!)^{\eta_i}}}{Dn\sum_{\{\eta_i\}}  \prod_i^{n}\frac{\psi_i^{\eta_i}}{\eta_i! (i!)^{\eta_i}}}.
\label{f(D) homo}
\end{equation} 
In all cases that we have been able to study to high $D$ (the meaning of `high' will be clarified later), $f_c(D) - f_c(\infty)$ is observed to scale as $1/D$ in the large $D$ limit (see figure \ref{1/D scaling}). The question of convergence speed then reduces to how large $D$ must be for this scaling to hold, and the the value of $f_c(D) - f_c(\infty)$ at this point. In general there are two distinct regimes of convergence, determined by the yield of target structures. We shall illustrate these regimes by considering completely cooperative transitions (in which only the target cluster and monomer concentrations are non-negligible), before commenting on the effects of other cluster sizes.

\subsection{Convergence at low yield}
Section \ref{homocluster formation} indicates that simulations of a single cluster underestimate the transition width and hence underestimate the yield of clusters at low yield. In effect, in order to have a high isolated monomer fraction in bulk despite the effects of volume fluctuations, the fraction of monomers in a single target simulation must be even higher. As the system size is increased, concentration fluctuations tend to transfer statistical weight from the extreme state favoured at $D=1$ towards a more balanced cluster size distribution.

\begin{figure}
\begin{center}
\includegraphics[width=7.2cm]{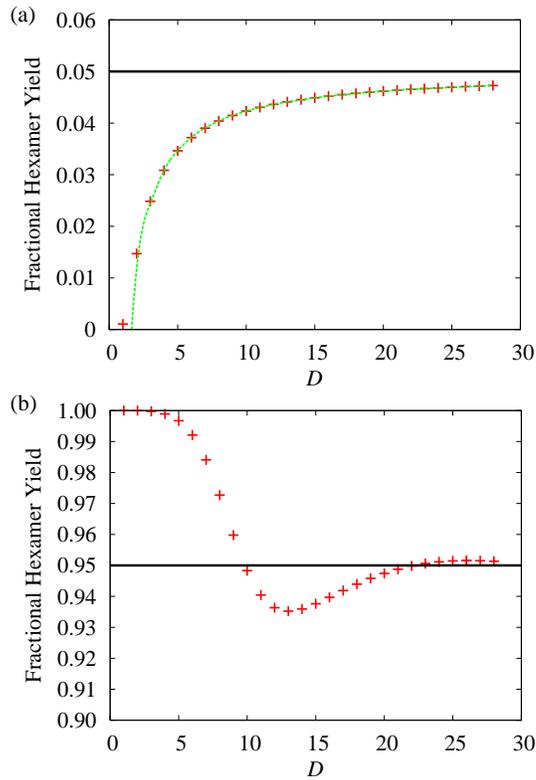}
\end{center}
\caption{\footnotesize Fractional yield of hexamers in a perfectly cooperative model as a function of system size $D$ for (a) low yield (5\% hexamers) and (b) high yield (95\% hexamers). The `+' symbols are the calculated points and the flat line the bulk value. The curve in (a) is a fit  to a $1/D$ convergence.}
\label{hexamers}
\end{figure}

At low yield this effect produces a steady increase in the proportion of clusters with $D$, with the deviation from the bulk fraction scaling as approximately $1/D$ from low $D$ (see figure \ref{hexamers}(a)). At very low yield, initial convergence becomes noticeably slower than $1/D$---this effect increases with target size. As a consequence, relative errors remain significant at increasingly large values of $D$ as the yield is decreased or the target size increased.

\subsection{Convergence at high yield}
At high yield, single target simulations overestimate the monomer fraction, for reasons similar to the underestimate at low yield. Convergence, however, does not initially show a $1/D$ behaviour, as illustrated in figure \ref{hexamers}(b). Instead, a period of slow convergence is followed by a rapid drop to a target yield just below the bulk value, leading eventually to an oscillation in the vicinity of the bulk yield. These oscillations persist for approximately $n-1$ half-cycles, before settling in to a $1/D$ convergence ($n$ being the target cluster size).  

These oscillations result from certain configurations disproportionately biasing the ensemble, due to the inherently discrete nature of a small system. At $D=1$, the system is restricted to the two states of one cluster or $n$ monomers. At high cluster yield, $n$ monomers are extremely unfavourable and hence the single cluster state is overwhelmingly observed, causing $f_n(1)$ to exceed $f_n(\infty)$. As the system size is increased, the zero monomer state continues to exert a disproportionate influence on the ensemble, keeping $f_n(D)$ well above $f_n(\infty)$. Eventually, however, the system becomes sufficiently large that the state with $D-1$ clusters is most favourable. Due to the discreteness of the system, this occurs before $(D-1)/D = f_n(\infty)$. As a consequence, $f_n(D)$ is then underestimated (or equivalently the number of monomers is overestimated), resulting in the observed drop of $f_n(D)$. At still larger values of $D$, the state with $D-1$ clusters remains most favourable but now constitutes an over estimate of $f_n(D)$, resulting in the observed rise in $f_n(D)$. This process is repeated for increasing number of monomers, leading to oscillations which are eventually overwhelmed by the $1/D$ convergence at large system size.

\begin{figure}
\begin{center}
\includegraphics[width=7.2cm]{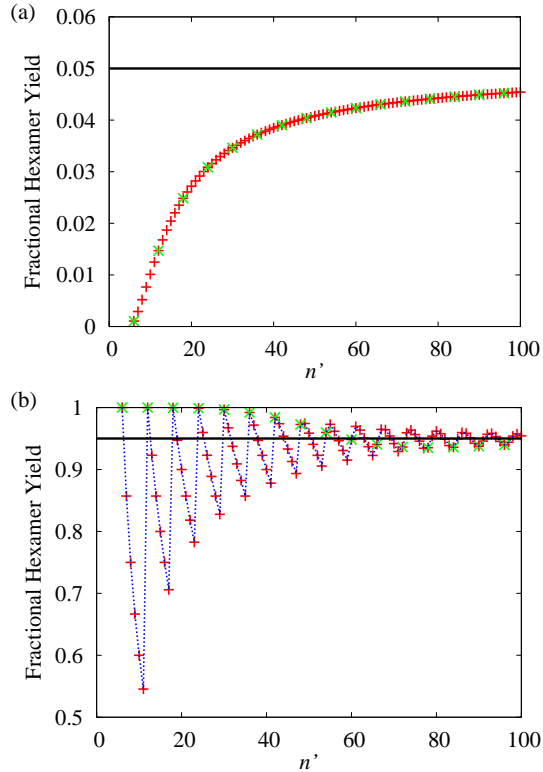}
\end{center}
\caption{\footnotesize Fractional yield of hexamers in a perfectly cooperative model as a function of $n'$ for (a) low yield (5\% hexamers) and (b) high yield (95\% hexamers). The `+' symbols are the calculated points and the flat line the bulk value. `x' symbols indicate system sizes for which $D=n/n'$ is integral.
The dashed line in (b) is added as a guide to the eye.}
\label{hexamers_2}
\end{figure}

The question is then why oscillations are observed at high but not low yield, where the discreteness of the system is still present. To answer this, it is illuminating to allow $D$ to take non-integer values so that the system size $n'=Dn$ can take any integer value. At high yield, figure \ref{hexamers_2}(b), we see that the system is extremely sensitive to the exact number of particles, because if $D$ is not an integer there are necessarily excess monomers. This results in the rapid oscillation of $f_n(n')$ with a period of approximately $n$. Closer inspection, however, reveals that the period is longer than $n$, due to the fact that states with no monomers present become increasingly unfavourable as $D$ gets larger. The region in which the $f_n(n')$ peaks transfer from $n'  \bmod{ n} = 0$ to $n'  \bmod{ n} = 1$ corresponds to the region in which $f_n(D)$ drops off rapidly. By contrast, $f_n(n')$ increases monotonically with $n'$ at low yield (figure \ref{hexamers_2}(a)). 
In this regime, the fraction of clusters is not high enough for the value of $n'  \bmod{ n}$ to be significant, and so the general tendency to transfer statistical weight to states with a greater mix of cluster sizes is dominant, and smooth convergence is observed.  

As a consequence of this behaviour, convergence at high cluster yield is extremely poor until $D$ is sufficiently large that the state with $D-1$ clusters has approximately the same weight as the state with $D$ clusters:
\begin{equation}
\frac{(\psi_n)^{D-1} \psi_1}{(D-1)! (n!)^{D-1} n!} \approx \frac{(\psi_n)^{D} }{(D)! (n!)^{D}}.
\end{equation}
Substituting using the definition of $\psi_i(D)$ gives:
\begin{equation}
D_{\rm crosssover} \approx \left(\frac{z_n}{z_1^n} \right)^{1/n}.
\end{equation}
The quantity $z_n/z_1^n$ corresponds to the ratio of cluster to monomers at $D=1$, and consequentially increases with $n$ at fixed bulk yield. This increase is offset by the $1/n$ exponent, meaning that the value of $D_{\rm crosssover}$ 
is relatively independent of target size, but increases with the target yield. It should also be noted that the oscillations persist until a system size of approximately $nD_{\rm crossover}$, although they are generally reasonably small. It is this value, $D \approx nD_{\rm crossover}$, that defines the large $D$ limit.

In the intermediate yield regime near to the midpoint of the transition, the initial error is small and $z_n/z_1^n$ is not large, hence convergence is fast (whether it proceeds by the first or second method). Away from the midpoint, however, significant relative discrepancies can persist to surprisingly large system sizes.  

\subsection{Intermediate cluster sizes}
\begin{figure}
\begin{center}
\includegraphics[width=7.2cm]{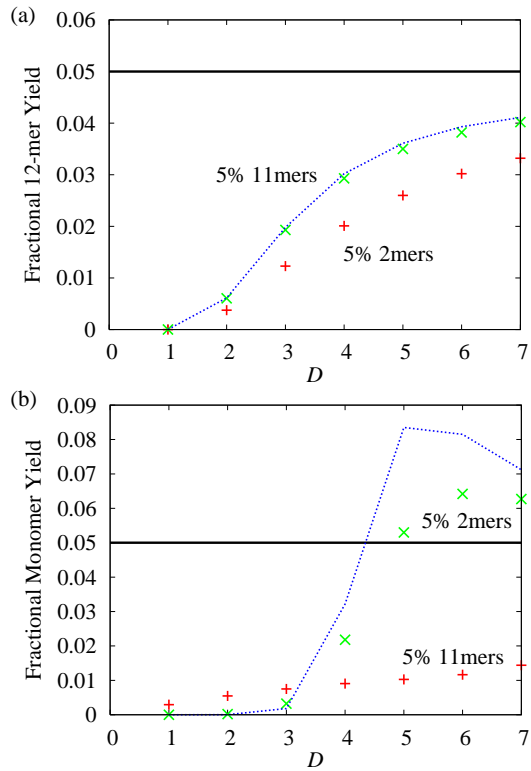}
\end{center}
\caption{\footnotesize (a) Fractional yield of dodecamers in a dodecamer forming system at low yield (90\% isolated monomers, 5\% dodecamers in bulk). (b) fractional yield of isolated monomers in a dodecamer forming system at high yield (90\% dodecamers, 5\% monomers in bulk). Plotted are points for systems in which the other 5\% is assumed to consist entirely of either 2-mers or 11-mers. 
Also shown (dashed curves) are the results for completely cooperative systems with the same 90\% majority species yield. These have been scaled by a factor of 0.5 so that the relative errors can be directly compared.}
\label{dodecamer}
\end{figure}

An additional complication for $n > 2$ is the fact that intermediate cluster sizes may be relevant to the system, which can complicate convergence. We shall analyze the effects of the presence of intermediate cluster sizes under the assumption that the majority of particles are found either as isolated monomers or in the target cluster size: for the purposes of this section, the term `majority species' applies to the most prevalent of either the target cluster or monomers, and `minority species' to the less common of these two. Note that our discussions will compare the effects of intermediate cluster sizes in systems with a certain yield of the majority species, as it is the tendency of one species to dominate in bulk despite concentration fluctuations that causes the large discrepancies at $D=1$. Firstly, we shall consider the low yield case. Here, the presence of clusters of intermediate sizes with bulk yields comparable to the target cluster has little effect on the relative error of the target yield at $D=1$, which is largely determined by the bulk fraction of monomers. By contrast, if the relevant intermediate cluster size is small (for instance a dimer in a system forming a dodecahedron), the relative error between dimer and isolated monomers is comparatively small, meaning that $f_2(1) \approx f_2(\infty)$, because from the perspective of the monomer/dimer equilibrium the system has an effective size of $D_{\rm eff} = n/2$. As a consequence, states including dimers are common and so the entropic penalty associated with having no target clusters is reduced, meaning that statistical weight is transferred to larger clusters more slowly as the system size is increased. 
The effect manifests itself as a poor convergence in the first few steps, as shown in figure \ref{dodecamer}(a). Also shown is the effect of having a significant presence of large intermediate clusters, which is smaller as they do not relieve the entropic penalty of having many monomers as swiftly as dimers do (the relative error is seen to behave similarly to a completely cooperative system with the same monomer yield).

We now consider the effect of significant presence of intermediate clusters on the convergence of the yield of {\em isolated monomers} at high cluster fraction. If the relevant intermediate clusters are large, the initial error is significantly reduced as the relative error between two large clusters of similar size is much smaller than for a large cluster and a monomer, and in forming intermediate clusters some monomers are `spare'. Convergence, however, is not improved as instead of the state with $D-1$ target clusters and $D$ monomers coming to dominate the ensemble, as in the completely cooperative case, states containing intermediate clusters become most prevalent (in effect, they reduce the `entropy cost' associated with having few monomers in the system). If the intermediate clusters are large, there will be few monomers in these states and as a consequence, statistical weight is transferred to isolated monomers more slowly. 
This effect is illustrated in figure \ref{dodecamer}(b): also plotted is a case with a significant presence of small intermediate clusters. In this case convergence is not dramatically slowed (relative to a completely cooperative system with the same target cluster yield), as the states which become prevalent contain $D-1$ target clusters and a mix of smaller species, including several monomers.

In summary, for monodisperse clusters, the significant presence of intermediate cluster sizes tends to reduce the rate of convergence of the fractional yield of the minority species relative to a completely cooperative system (at a fixed yield of the majority species), particularly if the relevant intermediate clusters are closer in size to the majority species, by reducing the entropic penalty associated with having few of the minority species in the system. 
In several cases, authors have studied systems capable of forming approximately 10--20 clusters \cite{Hagan2006, Nguyen2007, Berger1998}. 
It is probable that the finite-size effects illustrated here are relevant to these systems in the regimes dominated by one cluster size. 

\begin{figure}
\begin{center}
\includegraphics[width=5.2cm, angle=-90]{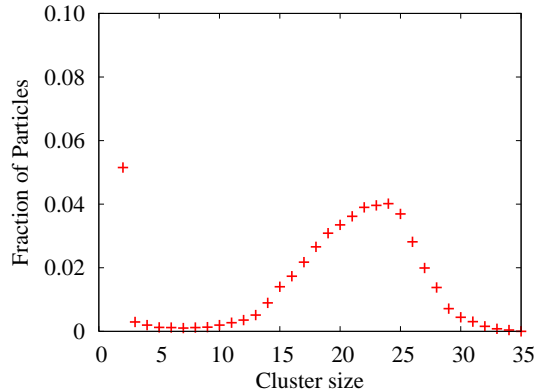}
\caption{\footnotesize Fractional number of particles in each cluster size at the CMC for the statistical micelle model used in the text. Here we define the CMC as the point at which half of all particles are in clusters larger than one (thus the fraction of monomers is 0.5).}
\label{micelle distribution}
\end{center}
\end{figure}

\subsection{Polydisperse large homoclusters}
In previous sections we have focused on monodisperse clusters, in which we have assumed that the majority of particles are either isolated monomers or in clusters of a certain size. We now consider self-assembly of structures which, although characteristically finite, have a much larger range of sizes. In particular, many simulations have studied the formation of spherical micelles from surfactants \cite{Viduna98, Gottberg1997, Milchev2001, Zehl2006}. We apply the theory developed earlier 
to estimate finite size statistical corrections for a model system whose bulk distribution is reasonably reflective of models in the literature. For the purposes of this investigation, we specify a cluster size distribution (in the bulk limit) at the critical micelle concentration (CMC), as shown in figure \ref{micelle distribution}, from which we infer $z_i({\rm CMC})$. Assuming all other factors are held constant, we then adjust the total concentration, adjusting $z_i$ accordingly, and observe the convergence of cluster yields on bulk values. 

\begin{figure}
\begin{center}
\includegraphics[width=7.2cm]{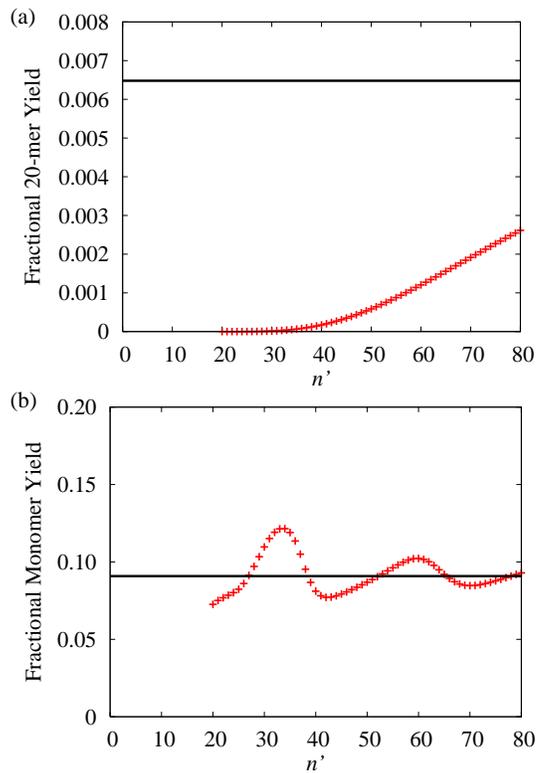}
\end{center}
\caption{\footnotesize (a) Fractional yield of micelles of size 20 in the low yield case (when approximately 10\% of particles are in micelles). (b) Fractional yield of isolated monomers in the regime where yield of micelles is high (90\%). The `+' symbols are the calculated points and the flat line the bulk value}
\label{micelle}
\end{figure}

Figure \ref{micelle}(a) shows the convergence of  a typical micellar cluster 
(20-mers) to its bulk yield, at a concentration at which approximately 10\% of particles are in micelles. In this case there is no single target structure size, so we plot the yield as a function of $n'$. 
In this regime, low target yield, the behaviour is very similar to that of monodisperse structures, with the fraction of 20-mers eventually converging toward the bulk value (with a limiting form of $1/n'$). At this concentration, the second most populated cluster is a dimer, and hence the initial convergence of the 20-mers is slowed as discussed in the previous section---even at a system size capable of forming four micelles, the fraction is less than half of its bulk value.

Figure \ref{micelle}(b) displays the convergence of isolated monomer fraction at a concentration when approximately 90\% of particles are in micellar structures. Although a large majority of the particles in the polydisperse system are in micellar structures, the polydispersity reduces the strength of the oscillations (note: these are oscillations in $f_i(n')$, the equivalent of those shown in figure \ref{hexamers_2}(b), not the gentler oscillations in $f_i(D)$ as exemplified in figure \ref{hexamers}(b)) as each different micelle size tends to oscillate out of phase. In addition, the large variety of micellar sizes means that no single cluster has a large fractional yield, meaning that the initial corrections are considerably smaller than in the monodisperse case. As a result, convergence is more successful for polydisperse than monodisperse clusters at high yield.

For micelles, therefore, errors are most likely to be significant at concentrations slightly below the critical micelle concentration, particularly when dimers are the second most common aggregate. Our model suggests that these errors may persist for systems containing several times the typical aggregation number of monomers, as large as some simulated systems \cite{Viduna98, Gottberg1997, Milchev2001, Zehl2006}, possibly affecting the details of cluster size distributions. 

\section{Variable dimensional umbrella sampling}
Given the complexity of corrections to simulations of small systems in the canonical ensemble, it would be beneficial to avoid the necessity of using them, particularly in systems where the ideality assumption is questionable, for example, if the interactions are long ranged with respect to the typical separation of 
unbound monomers. 
It would therefore be useful to have methods that can force the reversible
formation of multiple targets.

Conventional umbrella sampling enhances sampling by incorporating a biasing potential into the simulation which encourages the system to pass through transition states that would otherwise be improbable. The biasing potential is a function of a fixed number of `reaction coordinates': generally collective coordinates which describe the pathway of a transition.  We propose extending the umbrella sampling technique to a variable number of dimensions to allow formation of several targets. 

Let us assume that we have an unambiguous way of defining which monomers in our system constitute a cluster. In this case, we can define a biasing potential $u_{\rm bias}(n, {\bf x}^n)$, which is a function of the number of monomers in the cluster, $n$, and the coordinates of those monomers, ${\bf x}^n$. We then introduce the total biasing potential,
\begin{equation}
U_{\rm bias}({\bf x}^N) = \sum_i  u_{\rm bias}(n_i, {\bf x_i}^{n_i}),
\end{equation}
where the sum is over all clusters and 
which is used to bias sampling according to the factor $\exp(-\beta U_{\rm bias}({\bf x}^N))$. $u_{\rm bias}(n, {\bf x}^n)$ can be chosen by iteration analogously to conventional umbrella sampling, in order to favour transition states for each cluster. As the number of clusters in a given simulation is variable, the dimension of the effective `reaction coordinate' is variable. For this reason it is not appropriate to record states in a histogram according to their reaction coordinate, and so unbiasing must be performed on-the-fly.

\begin{figure}
\begin{center}
\includegraphics[width=7.2cm]{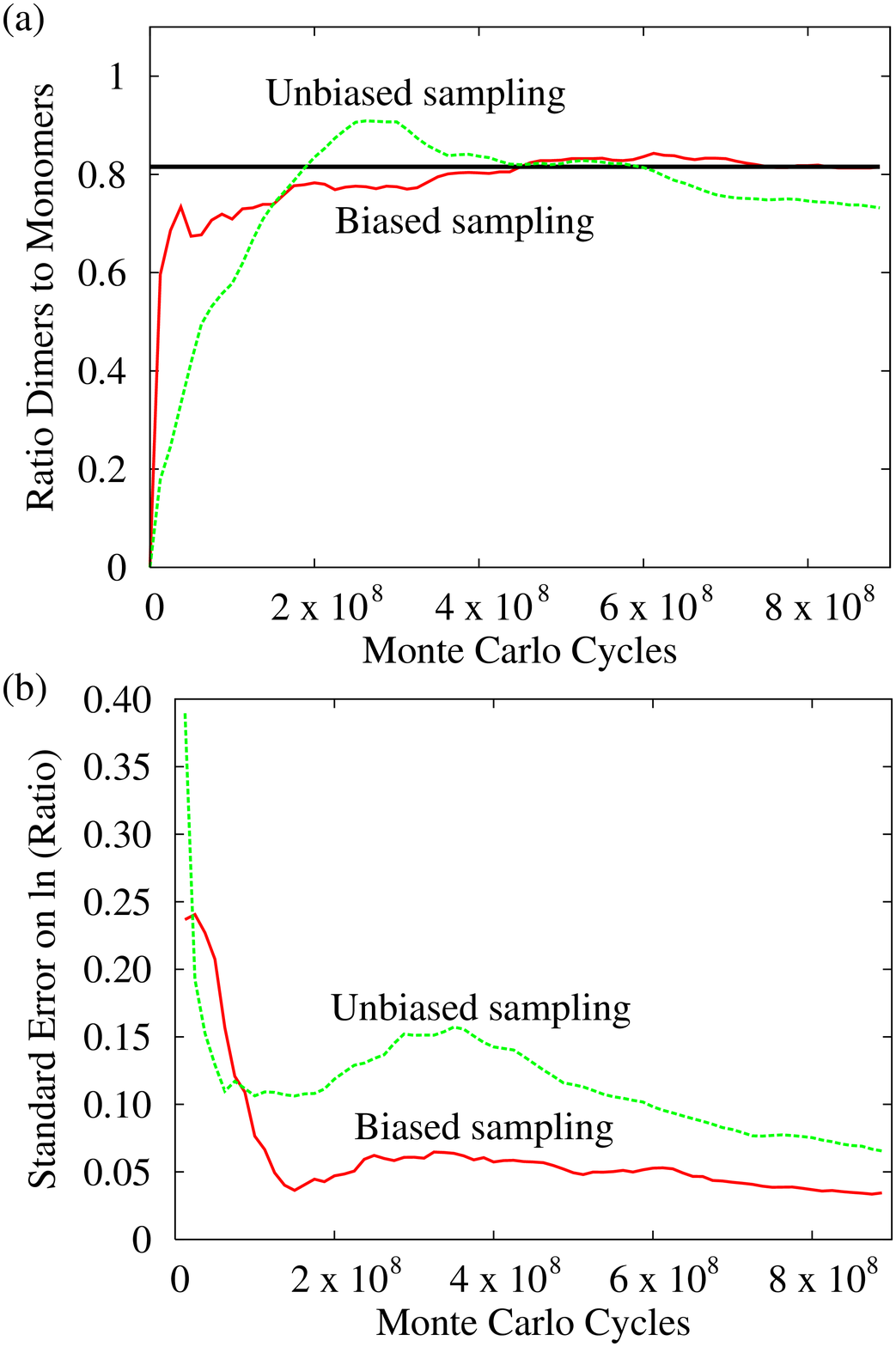}
\end{center}
\caption{\label{variable dimensional umbrella} Convergence of (a) the average 
ratio of dimers to monomers and (b) the error in the logarithm of this ratio 
in simulations of DNA self-assembly. 
Each simulation was performed on a system of eight strands each eight bases long
and the results are averaged over 10 independent simulations.
The results from simulations using the variable dimensional umbrella sampling scheme (biased) are compared to standard unbiased simulations. 
The flat line in (a) indicates the true equilibrium value 
(calculated through additional independent simulations). }
\end{figure}

To demonstrate the utility of the variable dimensional umbrella sampling scheme, we apply it to the simulation of the DNA model used in section \ref{heterodimer microscopic}. For a system of eight strands, each of eight bases (four strands of one type and four of its complement), we compare the convergence of the estimated bonding fraction in unbiased canonical simulations to that using variable dimensional umbrella sampling with a $u_{\rm bias}(n, {\bf x}^n)$ optimized by hand (in this case $u_{\rm bias}$ was simply a function of the number of base pair contacts). The results, shown in figure \ref{variable dimensional umbrella}, indicate a significant improvement in convergence over unbiased sampling, and
this advantage is likely to increase significantly for longer DNA strands.

Umbrella sampling, by its nature, increases the sampling of states which are not strongly represented in the ensemble---in the case of our DNA model, this corresponds to partially bound double strands. In conventional umbrella sampling, biasing potentials are often chosen to sample all points along the reaction coordinate with equal probability. If $u_{\rm bias}(n,{\bf x}^n)$ is chosen the same way, applying the scheme to a large system will result in the vast majority of states sampled involving several intermediate states: physically relevant states will be sampled rarely. To sample the physical states, one must reduce the biasing of the intermediate states, which seems to contradict somewhat the original 
purpose of the umbrella potential.

For a small system, however, a compromise is possible. $u_{\rm bias}(n,{\bf x}^n)$ can be chosen to favour intermediate states, but not so strongly that they are as likely as the physically relevant states. 
If this is done, transitions between the physically important states will be accelerated and they will also be sampled with reasonable frequency.
As such, an appropriate use of the technique would be on a small system in the grand canonical ensemble, similar to the study of Pool and Bolhuis \cite{Pool2005}, but in this case encouraging the formation of multiple clusters. If the simulation volume contains approximately one cluster on average, the total number of clusters simulated will remain small enough that the physically relevant states can be sampled frequently and efficiently, whilst also providing statistics which are directly applicable to bulk.

We should note that, although this variable dimensional umbrella sampling 
scheme can accelerate the sampling of systems with multiple targets, it is not 
practical to use it with umbrella potentials which require several simulation 
windows. 
This restriction thus limits the complexity of the system to which it can 
be applied.

\section{Conclusions}
We have demonstrated the cause of deviations from bulk statistics in finite size simulations in the canonical ensemble, and devised a method for estimating and correcting them under the assumption that species behave ideally. As simulation size is increased, fractional yields are found to converge in qualitatively different ways depending on the type of cluster which is most prevalent in the system: in general converging in a smoother fashion at low target structure yield, and oscillating at high yield. We also find that the discrepancies increase with distance from the midpoint of the transition. 

This study has highlighted a particular reason to be wary of statistical finite size effects.
As the convergence of abundances to their bulk values is strongly dependent on the yield of clusters, it is not sufficient to estimate finite size effects at one set of conditions and assume they apply at another---all regimes of interest must be checked. 

In answer to our original question of how best to compute the thermodynamics
of systems that self-assemble into finite-sized objects, 
where possible, we recommend performing simulations in the grand canonical 
ensemble (where the correct concentration fluctuations are naturally generated)
or using a system large enough for the errors associated with the canonical
ensemble (as estimated by our prescription) to be negligible. 
However, such an approach requires that the reversible formation of multiple 
assembled structures is feasible on the available computational time scales.
For instances where this is not the case due to the large free energy barriers 
associated with assembly, an efficient alternative is to utilize a rare-event 
method such as umbrella sampling to simulate the assembly of a single target 
in the canonical ensemble, and then to apply the corrections outlined in the
current paper to obtain the bulk yield. For example, this is the approach we 
have taken for the numerous melting point calculations that were required in
the development and testing of our recent coarse-grained model for DNA \cite{Ouldridge_tw_2009}.

Finally, the proposed extension to the umbrella sampling scheme which 
individually weights each cluster as it appears in the simulation has been
shown to aid the reversible formation of multiple target structures. 
This scheme could be used to test the approximations of ideality inherent in 
the finite-size corrections by performing canonical simulations where increasing numbers of target structures can be formed. It could also be particularly useful 
to aid sampling in grand canonical simulations in cases where free energy
barriers make equilibrium otherwise difficult to achieve.


\end{document}